\documentclass[fleqn,usenatbib]{mnras}

\usepackage{amsmath}
\usepackage{txfonts}                
\usepackage{graphicx}               
\usepackage{ragged2e}

\title[Prolate-like rotation in massive galaxies]{Climbing to the top of the galactic mass ladder: evidence for frequent prolate-like rotation among the most massive galaxies}

\author[Davor Krajnovi\'c et al.]{Davor Krajnovi\'c $^{1}$\thanks{E-mail:dkrajnovic@aip.de}, Eric Emsellem$^{2,3}$, Mark den Brok$^{4,1}$, 
Raffaella Anna Marino$^{4}$,  \newauthor Kasper Borello Schmidt$^{1}$, Matthias Steinmetz$^{1}$ and Peter M. Weilbacher$^{1}$\\
\\
$^{1}$Leibniz-Institut f\"ur Astrophysik Potsdam (AIP), An der Sternwarte 16, D-14482 Potsdam, Germany\\
$^{2}$ESO, European Southern Observatory, Karl-Schwarzschild Str. 2, 85748 Garching bei Muenchen, Germany\\
$^{3}$Univ Lyon, Univ Lyon 1, ENS de Lyon, CNRS, Centre de Recherche Astrophysique de Lyon UMR 5574, 69230 Saint-Genis Laval, France\\
$^{4}$ETH Zurich, Department of Physics, Wolfgang-Pauli-Str. 27, 8093 Zurich, Switzerland\\
}

\date{Accepted 2018 April 19. Received 2018 April 19; in original form 2018 February 7}

\pubyear{2017}

\begin{document}
\label{firstpage}
\pagerange{\pageref{firstpage}--\pageref{lastpage}}
\maketitle

\begin{abstract}
We present the stellar velocity maps of 25 massive early-type galaxies located in dense environments observed with MUSE. Galaxies are selected to be brighter than M$_K=-25.7$ magnitude, reside in the core of the Shapley Super Cluster or be the brightest galaxy in clusters richer than the Virgo Cluster. We thus targeted galaxies more massive than 10$^{12}$ M$_\odot$ and larger than 10 kpc (half-light radius). The velocity maps show a large variety of kinematic features: oblate-like regular rotation, kinematically distinct cores and various types of non-regular rotation. The kinematic misalignment angles show that massive galaxies can be divided into two categories: those with small or negligible misalignment, and those with misalignment consistent with being 90\degr. Galaxies in this latter group, comprising just under half of our galaxies, have prolate-like rotation (rotation around the major axis). Among the brightest cluster galaxies the incidence of prolate-like rotation is 50 per cent, while for a magnitude limited sub-sample of objects within the Shapley Super Cluster (mostly satellites), 35 per cent of galaxies show prolate-like rotation. Placing our galaxies on the mass - size diagram, we show that they all fall on a branch extending almost an order of magnitude in mass and a factor of 5 in size from the massive end of galaxies, previously recognised as associated with major dissipation-less mergers. The presence of galaxies with complex kinematics and, particularly, prolate-like rotators suggests, according to current numerical simulations, that the most massive galaxies grow predominantly through dissipation-less equal-mass mergers.

\end{abstract}




%
%

\section{Introduction}
\label{s:intro}

The orbital structure is a powerful tracer of the formation processes shaping galaxies. As galaxies acquire gas, accrete satellites or merge with similar size objects, new populations of stars are created and the mass and luminosity distributions evolve. The changes in the gravitational potential have a direct influence on the allowed and realised trajectories, providing for a variety of observed stellar kinematics. As observers, we thus hope to constrain the ingredients (and chronology) which shaped galaxies by probing the spatial variations of the line-of-sight velocity distribution (LOSVD). 

Theoretical insights, based on analytical and numerical work, are crucial for the interpretation of the observed stellar kinematics of galaxies \citep[see e.g.,][]{1991ARA&A..29..239D}. In an idealised system with triaxial symmetry, assuming a gravitational potential expressed in a separable form \citep[e.g. St\"ackel potentials as introduced by][]{1915MNRAS..76...37E}, there exist a few families of dissipation-less orbits which stars can adopt: box orbits, short-axis tubes, inner and outer long-axis tubes \citep{1985MNRAS.216..273D}. In such systems, symmetry changes, for example between spherical, oblate or prolate axial symmetries, limit the stability of orbital families. \citet{1985MNRAS.216..273D} showed that a purely oblate spheroid should consist of only short-axis tubes, and therefore show a typical streaming {\it around} its minor axis, unless there is an equal amount of stars on both prograde and retrograde orbits canceling out the net streaming. A prolate spheroid allows only inner and outer long-axis tubes, and streaming {\it around} the major axis of the galaxy. The argument can also be reversed to state that galaxies with only long axis tubes cannot be oblate and axisymmetric, or even triaxial, and that a galaxy with short axis tubes does not have prolate symmetry.

The velocity maps of triaxial spheroids, viewed at random angles, can exhibit a rich variety of kinematic features. This is a direct consequence, as pointed out by \citet{1991ARA&A..29..239D}, of the freedom in the direction of the total angular momentum resulting from the orbital mixture, and the momentum vector which can lie anywhere in the plane containing the major and minor axis of the galaxy. This was illustrated by \citet{1991AJ....102..882S} with models viewed along various orientation angles, and associated with actually observed galaxies with complex kinematics \citep[e.g. NGC\,4356 and NGC\,5813;][respectively]{2008MNRAS.385..647V, 2015MNRAS.452....2K}

Observational studies using long-slits were able to investigate velocity features along selected angles (often along the minor and major photometric axes), and revealed that a majority of galaxies exhibit negligible rotation along their minor photometric axis \citep[e.g.][]{1983ApJ...266...41D,1983ApJ...266..516D, 1994MNRAS.269..785B}, while a few massive elliptical galaxies show more complex rotation indicating the presence of long-axis tubes and significant rotation around their major axis \citep[e.g.][]{1977ApJ...218L..43I,1979ApJ...229..472S,1988A&A...195L...5W}. A major change in this field came from the proliferation of the integral-field spectrographs (IFS) and their ability to map the distribution of velocities over a significant fraction of the galaxy. The last decade of IFS observations has revealed that the vast majority of galaxies actually has very regular velocity maps within their half light radii \citep[e.g.][]{2004MNRAS.352..721E, 2011MNRAS.414.2923K, 2013MNRAS.436...19H, 2014MNRAS.441..274S, 2015MNRAS.454.2050F, 2017A&A...597A..48F,2018MNRAS.tmp..522G}. 

The ATLAS$^{\rm 3D}$ project \citep{2011MNRAS.413..813C} addressed this more specifically via a volume limited survey of nearby early-type galaxies, demonstrating that galaxies with complex velocity maps comprise only about 15\% of the local population of early-type galaxies \citep{2011MNRAS.414.2923K}, and that the majority is consistent with oblate rotators \citep[notwithstanding the presence of a bar, see][]{2011MNRAS.414.2923K, 2014MNRAS.444.3340W}. The regular and non-regular rotator classes seem to reflect a significant difference in their specific stellar angular momentum content, allowing an empirical division of early-type galaxies into fast and slow rotators \citep{2007MNRAS.379..401E,2011MNRAS.414..888E}. \citet{2008MNRAS.390...93K} also emphasised the fact that axisymmetric fast rotators have regular velocity fields which qualitatively resemble those of disks. The internal orbital structure of these galaxies can, however, be complex, as evidenced by the range of photometric properties (e.g. disk-to-bulge ratio) and the common presence of tumbling  bars.

There are several caveats which need to be emphasised. Firstly, the intrinsic shape of a galactic system is seldom well defined by a single number, e.g., the apparent ellipticity varies with radius. Along the same lines, the terms "triaxial" or "oblate" systems may not even be appropriate when the intrinsic ratios and/or the position angle of the symmetry axes change with distance from the centre: the gravitational potential of a galaxy could smoothly vary from oblate in the centre to strongly triaxial or prolate in the outer part, with the main symmetry axes not even keeping the same orientation. Secondly, ellipsoids are certainly a very rough approximation when it comes to describing the intrinsic shapes of galaxies, as they have overlapping components with different flattenings, varying bulge-to-disk ratios, and often host (tumbling) bars. While the observed kinematics of fast rotators \citep[including also higher moments of the LOSVD,][]{2008MNRAS.390...93K, 2011MNRAS.414.2923K, 2017ApJ...835..104V} seem to indicate that their internal orbital structure is dominated by short-axis tube orbits (and streaming around the minor axis), numerical simulations of idealised mergers and those performed within a cosmological context naturally predict the coexistence of multiple orbital families, the central and outer regions often being dominated by box and short-axis tube orbits, respectively \citep[e.g.][]{2005MNRAS.360.1185J,2010ApJ...723..818H,2014MNRAS.445.1065R}. 

The division of galaxies into fast and slow rotators connects also with two dominant channels of galaxy formation \citep[as reviewed in][]{2016ARA&A..54..597C}. Present spirals and fast rotators are mostly descendants of star forming disks and their evolution is dominated by gas accretion, star formation, bulge growth and eventual quenching. The slow rotators may also start as turbulent star-bursting disks at high redshift \citep[e.g.][]{2009ApJ...703..785D, 2009MNRAS.395..160K}, or be hosted by haloes with small spin parameters \citep{2017arXiv171201398L}, but the late evolution of most slow rotators is dominated by mergers with gas poor galaxies \citep{2007MNRAS.375....2D, 2009ApJ...703..785D, 2011ApJ...738L..25W, 2015MNRAS.452.2845K}. The first channel therefore favours regular kinematics and internal orbital structure dominated by short-axis tubes, while the second channel implies dynamically violent redistribution of orbits and the creation of triaxial or prolate-like systems, which include a significant fraction of long-axis tubes. A strong mass dependence has been emphasised, with more massive galaxies being more likely to follow the second channel \citep{2016MNRAS.458.2371R,2017MNRAS.464.1659Q}.

A clear manifestation of the triaxial nature of galaxies is a non-zero value of the kinematic misalignment angle, $\Psi$, the angle between the photometric minor axis and the orientation of the apparent angular momentum vector \citep[][]{1991ApJ...383..112F}. In an axisymmetric galaxy, the apparent angular moment coincides with the intrinsic angular momentum and is along the minor axis, hence $\Psi=0$. Triaxial galaxies can exhibit any value of $\Psi$, while prolate galaxies with significant rotation would have $\Psi$ closer to 90\degr. Galaxies with large $\Psi$ exist \citep{1991ApJ...383..112F,2007MNRAS.379..418C,2011MNRAS.414.2923K,2015MNRAS.454.2050F,2017A&A...606A..62T} and are typically more massive than $10^{11}$ M$_\odot$ \citep[but for dwarf galaxies see e.g.][]{2012ApJ...758..124H, 2013MNRAS.428.2980R}. It is, however, not clear if prolate-like systems feature prominently at high mass and if this links preferentially to a specific channel of galaxy evolution.

Galaxies at the top of the mass distribution are intrinsically rare. They are mostly found in dense environments, often as the brightest members of groups or clusters. Brightest cluster galaxies (BCGs) are usual suspects, and are known to have low amplitude or zero rotation \citep{2008MNRAS.391.1009L,2013ApJ...778..171J, 2017AJ....153...89O}. Still, current surveys of massive galaxies have so far offered little evidence for large $\Psi$ values or clear-cut signatures for strong triaxiality \citep[e.g.][]{2017MNRAS.464..356V}. In this work, we present the first results from an observation-based survey, the M3G (MUSE Most Massive Galaxies; PI: Emsellem) project, aimed at mapping the most massive galaxies in the densest galaxy environments at $z\approx0.045$ with the MUSE/VLT spectrograph \citep{2010SPIE.7735E..08B}. We focus on presenting the stellar velocity maps, emphasising the relatively large number of prolate-like systems, i.e., galaxies with rotation {\it around} the major axis. The orbital distribution of galaxies exhibiting large values of $\Psi$ (and having net rotation around the major axis) are thought to be dominated by long-axis tubes: we will thus refer to such cases as {\it prolate-like} rotation\footnote{An alternative name used in the literature is {\it minor-axis} rotation, as the gradient of the velocities is {\it along} the major axis, and it should be differentiated from rotation {\it around} the minor axis present in oblate axisymmetric systems. To avoid any ambiguity, next to the defined prolate-like rotation, we will use the nomenclature from \citet{2011MNRAS.414.2923K}, where regular rotation is used for oblate rotators with velocity maps consistent with those of discs, while non-regular rotation is used for twisted and complex velocity maps.}. However, as mentioned above, and discussed in Section~\ref{s:discs}, we caution the reader that this does not imply that these are prolate systems. Presenting the complete survey, its data products and subsequent results is beyond the scope of the current publication and will be presented in forthcoming papers. 

In Section~\ref{s:obs} we briefly report on the observations and the data analysis. We present the main results on the rotational characteristics of the M3G sample in Section~\ref{s:prolate}, which is followed by a discussion in Section~\ref{s:discs} and a brief summary of conclusions in Section~\ref{s:con}.

%
%

\section{Observations and analysis}
\label{s:obs}

In this section we briefly describe the M3G sample of galaxies, the observations and the extraction of the kinematic information. Further details on these aspects will be presented in a following M3G paper (Krajnovi\'c et al. in prep.). 

\begin{figure*}
\includegraphics[width=\textwidth]{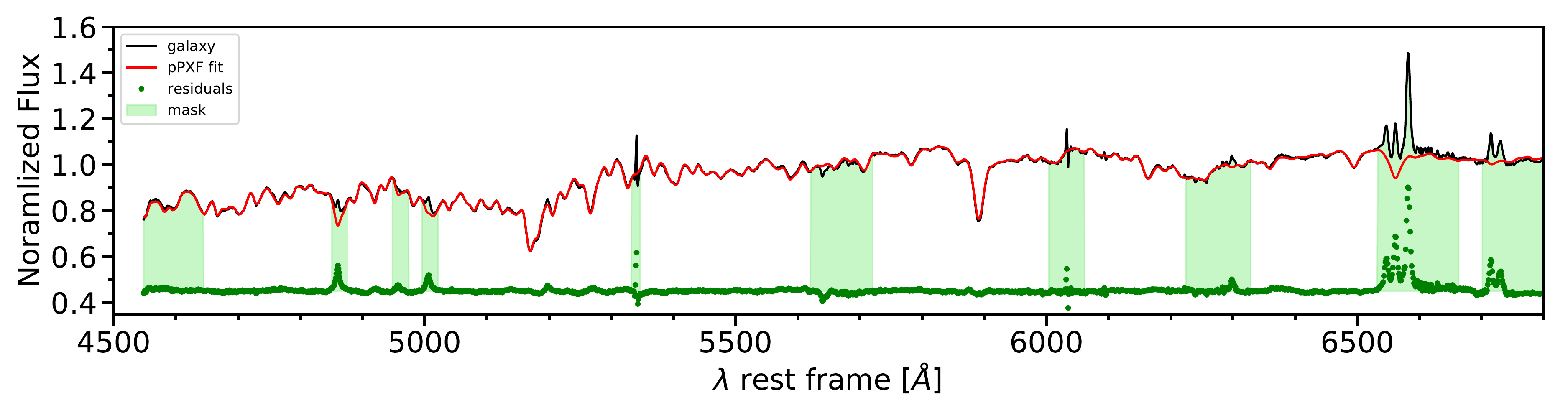}
\caption{An example of the pPXF fit to a spectrum extracted within an effective radius from PGC047177, which also shows ionised gas emission. The observed spectrum is shown in black and the best fit in red. Green dots are residuals. Light green shaded areas were masked and not fitted. These include the strongest emission-lines expected between 4500 \AA\, and 7000 \AA, as well as potential strong sky lines or telluric residuals.}
\label{f:ppxf}
\end{figure*}

\subsection{The M3G sample and MUSE observations}
\label{ss:sample}

The M3G sample comprises 25 early-type galaxies selected to be brighter than -25.7 magnitude in the 2MASS K$_s-$band and found in the densest environments. We created two sub-samples of galaxies: one consisting of the brightest galaxies in the densest known structure, the core of the Shapley Super Cluster (SSC) \citep{1930BHarO.874....9S,2010MNRAS.402..753M,2015MNRAS.446..803M}, and the other targeting BCGs in rich clusters. We selected galaxies in the SSC using the 2MASS All Sky Extended Source Catalog \citep[XSC,][]{2000AJ....119.2498J,2006AJ....131.1163S} centred on the three main clusters near the core of the SSC: Abell 3562, 3558 and 3556 \citep{1989ApJS...70....1A}. This selection yielded 14 galaxies, with 3 being BCGs. The complementary sub-sample of BCGs was defined using a parent sample of clusters richer than the Virgo Cluster and observed with the HST \citep{2003AJ....126.2717L}. We included 11 BCGs residing in clusters with richness larger than 40, where the richness is defined as the number of galaxies with magnitudes between $m_3$ and $m_{3}+2$ within an Abell radius of the cluster centre ($m_3$ is the magnitude of the third brightest cluster galaxy). Here we also used the information given in \citet{2003AJ....126.2717L}. The full M3G sample therefore consists of 14 galaxies in the SSC, and 14 BCG (three being in the SSC). In this paper we use 2MASS photometry as a reference, but as part of the M3G project, we have collected photometry from other imaging campaigns, which will be described in detail in future papers. 

In addition to the visibility requirement that the galaxies are observable from Paranal, we imposed a selection criterion based on the distance and size of the galaxies: these had to be such that the MUSE field-of-view covers up to two effective radii of each target. The effective radii were collected from the XSC catalog, using the {\tt k\_r\_eff} keyword. The most massive galaxies in the SSC have the right combination of parameters to satisfy this criterion, while the additional 11 BCGs were selected to be at similar redshifts. The galaxies span the redshift range $0.037 < z < 0.054$, with a mean of z=0.046. The redshift of the SSC is assumed to be 0.048 \citep{1987MNRAS.225..581M}. Adopting cosmology H$_0=70$ km s$^{-1}$ Mpc$^{-1}$, $\Omega_{\rm M}=0.3$, $\Omega_\Lambda = 0.7$, 1\arcsec~is 904~pc at the mean redshift of the sample, while this scales changes from 735 to 1050~pc between galaxies \citep{2006PASP..118.1711W}. 

The observations of the sample were performed within the MUSE Guaranteed Time Observations (GTO) during ESO Periods 94 - 99 (starting in the fall of 2014 and finishing in the spring of 2017). The observing strategy consisted of combining a short Observing Block (OB) of exposures during better-than-average seeing conditions ($<0.8$\arcsec) to map the central structures, and a set of OBs with longer exposure times to reach a sufficient signal-to-noise ratio (S/N) at two effective radii. The high spatial (short exposure time) resolution MUSE data will be presented in a forthcoming paper. The total exposure time for each galaxy varied from about 2 to 6 hours. The brightest galaxy in the sample (see Table~\ref{t:sample} for details) was mosaiced with $2\times2$ MUSE fields, each observed up to 6h. All individual OBs consisted of four on-target observations and two separate sky fields sandwiched between the on-target exposures. On-target observations were each time rotated by 90\degr\ and dithered in order to reduce the systematics introduced by the 24 MUSE spectrographs. 

\subsection{Data reduction and kinematics extraction}
\label{ss:data}

Data reduction was performed as the observations were completed. This means that several versions (from v1.2 to the latest v1.6) of the MUSE data reduction pipeline \citep{2014ASPC..485..451W} were used. Despite continued improvement of the reduction pipeline, given the brightness of the M3G sample, and the nature of the current study, the differences in the reductions do not affect the results and conclusions presented here. All reductions followed the standard MUSE steps, producing the master calibration files of the bias and flat fields, as well as providing the trace tables, wavelength calibration files and line-spread function for each slice. When available we also used twilight flats. Instrument geometry and astrometry files were provided by the GTO team for each observing run. These calibrations files, as well as the closest in time illumination flats obtained during the night, were applied to the on-target exposures. From separate sky fields we constructed the sky spectra which were associated with the closest in time on-target exposure, and from the observation of a standard star (for each night) we extracted the response function as well as an estimate of the telluric correction. These, together with the line-spread function (LSF) and the astrometric solution, were used during the science post-processing. The final data cubes were obtained by merging all individual exposures. As these were dithered and rotated, a precise alignment scheme was required. This was achieved using stars or unresolved sources, and for a few cases in which the MUSE field-of-view was devoid of such sources, using the surface brightness contours in the central regions. The final cubes have the standard MUSE spatial spaxel of 0.2\arcsec$\times$0.2\arcsec\, and a spectral sampling of 1.25 \AA\, per pixel.

As a first step before extraction of the kinematics, we proceeded to spatially bin each data cube to homogenise the signal-to-noise ratio throughout the field-of-view via the Voronoi binning method \citep{2003MNRAS.342..345C}\footnote{\label{ft:CapSoft}Available at \href{http://purl.org/cappellari/software}{http://purl.org/cappellari/software}}. We first estimated the S/N of individual spectra from the reduction pipeline propagated noise, masking all stars or satellite galaxies within the field-of-view. Spatial binning is ultimately an iterative process, in which our goal was to achieve relatively small bins beyond one effective radius, but which provide a sufficient signal for extraction of robust kinematics. The quality of the extraction was measured using the signal-to-residual noise ratio (S/rN), where the residual noise is the standard deviation of the difference between the data and the model (as explained below). S/rN was required to be similar to the target S/N in bins at large radii. As the data quality varies between galaxies, it is possible for some galaxies to have sufficiently small bins in the central regions with S/rN $\sim100$, while for some galaxies S/rN $\sim50$ is the most that can be achieved for a reasonable bin size. For this work, we set the target S/N required by the Voronoi binning method to 50 for all galaxies. Additionally, before binning we removed all spectra (individual spaxels) with S/N less than 2.5 in the continuum (based on the pipeline estimated noise)\footnote{The value of S/N $\sim2.5$ was selected as a compromise between removing too many pixels in the outer regions and reducing the size of the outermost bins.}. In this way we excluded the spectra at the edge of the MUSE FoV, which essentially do not contain any useful signal and limited the sizes of the outermost bins. 

Stellar kinematics were extracted using the pPXF method\footnote{See footnote~\ref{ft:CapSoft} for software availability.} \citep{2004PASP..116..138C}. Our pPXF set up included an additive polynomial of the 4th order, and we fitted a line-of-sight velocity distribution parametrised by Gauss-Hermit polynomials \citep{1993ApJ...407..525V, 1993MNRAS.265..213G} with the mean velocity $V$, the velocity dispersion $\sigma$ and the higher order moments $h_3$ and $h_4$. We masked all potential emission-lines and a few narrow spectral windows with possible sky-line residuals. Finally, we limited the fit to blue-wards of 7000 \AA, to exclude potentially strong telluric and sky residuals. For each galaxy, a pPXF fit was first performed on the spectrum obtained by summing all MUSE spectra within one effective radius (covering an elliptical area equivalent to $\pi\times R_e^2$) and using the full MILES stellar library \citep{2006MNRAS.371..703S, 2011A&A...532A..95F} as templates. The MUSE LSF significantly varies with wavelength with a full-width half maximum from 2.85 \AA\, at 5000\AA\, to 2.5 \AA\, at 7000 \AA\, \citep{2017A&A...608A...5G}. We used the parametrisation of the LSF from \citet{2017A&A...608A...5G} and convolved the MILES templates to the MUSE LSF (varying with wavelength). Possible emission-lines were masked. As an example, we show the fit to the spectrum extracted within the half-light redius of one of our galaxies in Fig.~\ref{f:ppxf}. 

\begin{figure*}
\includegraphics[width=\textwidth]{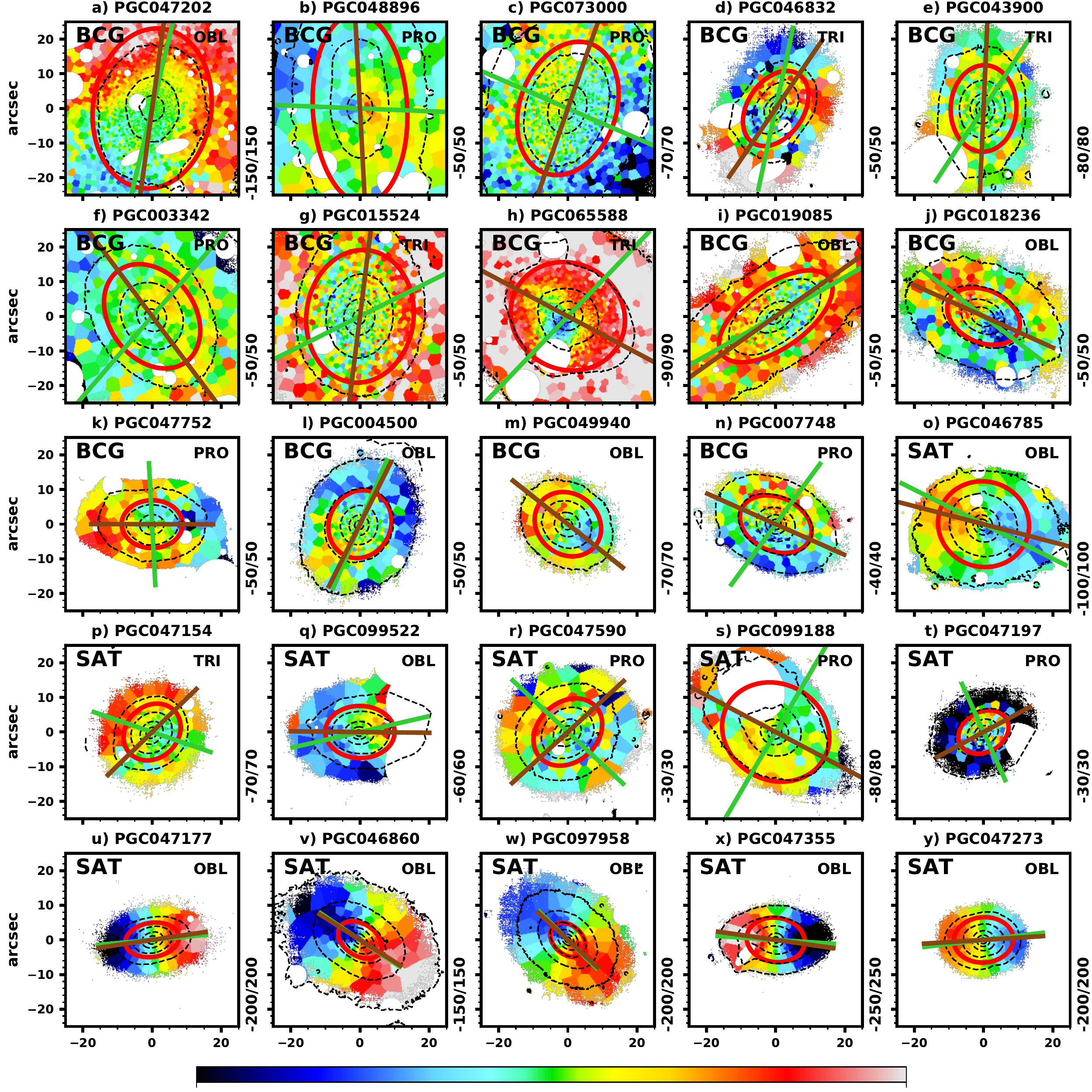}
\caption{The mean stellar velocity maps of the M3G sample galaxies. Galaxies are divided in two groups, the BCGs and satellites (non-BCGs in the SSC). BCGs are plotted in the first 14 panels starting from the top left, followed by satellites (as indicated with ``SAT"). The two groups of galaxies are ordered by decreasing $K$-band absolute magnitude. The values in the lower right corner of each panel indicate the range of the velocities, where the negative are shown with blue and positive with red colours, as indicated by the colourbar. Black dashed contours are isophotes plotted in steps of one magnitude. All velocity maps are approximately $1^{\prime}\times1^{\prime}$ in size. Full red ellipses indicate the size and the orientation of the half-light region, specified by the ellipticity of the galaxy and the semi-major axis length equal to the 2MASS K$s$-band effective radius. Green and brown lines indicate the orientation of the kinematic and the photometric major axes, respectively. Letters in upper right corner of each panel (``PRO", ``TRI" and ``OBL") indicate broad shape-related categories of the galaxy based on the kinematic misalignment (see Fig.~\ref{f:kma} for details).  Note that PGC\,043900 is characterised as "TRI" due to its non-rotation. The letters in front of the galaxy names will be used in text for easier location of the object.}
\label{f:maps}
\end{figure*}

This first global pPXF fit provides an optimal set of stellar templates, which we propagate for each individual Voronoi-binned spectrum, using the same pPXF set-up. The quality of the fit was checked via the S/rN of each bin, where the residual Noise was the standard deviation of the difference between the data and the best-fit pPXF model. As outlined above, this S/rN was required to be at least 50 over most of the field-of-view. We extracted up to the 4th Gauss-Hermit coefficient, but in this work we will primarily focus on the mean velocity maps of our 25 galaxies. The velocity dispersion and higher order velocity moments maps, as well as the analysis of the angular momentum will be presented in a future paper. The global velocity dispersion values are given in Table~\ref{t:sample}.

%
%

\section{Prevalence of long-axis rotation in massive galaxies}
\label{s:prolate}

The velocity maps for the full M3G sample are shown in Fig.~\ref{f:maps}. The sample is split into BCGs (first 14 maps) and non-BCGs from the SSC. In each subgroup galaxies are plotted in order of decreasing 2MASS $K$-band absolute luminosity. There are several noteworthy features in these maps, which we interpret within the kinematic classification system of the ATLAS$^{\rm 3D}$ survey \citep{2011MNRAS.414.2923K}. To start with, we note that almost all galaxies show some level of rotation. While the maximum velocity amplitudes reached within the two effective radii covered by our MUSE observations are often low ($\approx30 -50$ km/s), only one galaxy, {\it e)} PGC\,043900, does not show a clear indication for net streaming motion within the field of view. This is somewhat different from the trend expected from the ATLAS$^{\rm 3D}$ data \citep{2011MNRAS.414..888E,2011MNRAS.414.2923K}, where a few of the most massive systems (about 15 per cent for galaxies more massive than $2\times10^{11}$ M$_\odot$), can be characterised as having no net rotation. Other studies of massive galaxies \citep[e.g.][]{2017MNRAS.464..356V} also find a large number of galaxies with negligible net rotation. It is likely that, as in the case of NGC\,4486 \citep{2014MNRAS.445L..79E}, our MUSE data are of such quality and extent that the rotation is revealed even in systems such as {\it r)} PGC\,047590, where the amplitude of the rotation is only 30 km/s\footnote{A similar change of kinematic classification based on higher quality kinematics could happen to NGC\,5846, another ATLAS$^{\rm 3D}$ "non-rotator" with a hint for prolate-like rotation in the SAURON data.}. 

The coverage beyond one effective radius helps to determine the net rotation trend, but also reveals changes in the kinematics. This is especially noticeable among BCGs, where the velocity maps change orientation (e.g. {\it b)} PGC\,048896), or there is a loss of coherent motions (e.g. {\it h)} PGC\,065588). Non-BCGs, which we will call satellites in this context, do not show such changes. It might be the case that the changes are found at larger radii \citep[as for some lower-mass fast rotators,][]{2014ApJ...791...80A}, but there is no clear evidence for this within 2 R$_e$.

Another striking feature is that there are galaxies which show regular rotation, with peak velocity in excess of 200 km/s. These galaxies are in fact among the lower luminosity bin of our set of massive galaxies, and found within the group of satellites. Galaxies that belong to this class are {\it v)} PGC046860, {\it u)} PGC047177, {\it y)} PGC047273, {\it x)} PGC047355 and {\it w)} PGC097958. Their dynamical masses (see Section~\ref{s:discs}) are around $10^{12}$ M$_\odot$, and they are all among the most massive galaxies with regular rotation. Their existence is expected \citep[e.g.][]{2007MNRAS.378.1507B,2008MNRAS.391.1009L, 2017MNRAS.464..356V,2017arXiv171201398L}, although their number likely decreases with increasing mass \citep[e.g.][]{2011MNRAS.414.2923K, 2013ApJ...778..171J, 2013MNRAS.436...19H, 2017MNRAS.464..356V, 2017arXiv170401169B}. The fact that these galaxies are not found among BGCs is indicative of their less violent evolution maintaining the regular rotation. However, there is also the case of {\it a)} PGC\,047202, the largest and the most luminous galaxy in the SSC, and a BCG, which shows high level of rotation, albeit non-regular.  

Non-regular rotation is the most common characteristics of the M3G velocity maps. It is especially among BCGs, but it also occurs in non-BCGs. The existence of kinematically distinct cores (KDC), counter-rotation, the radial variation of the kinematic position angle, as well as the analysis of the velocity features beyond the effective radius will be discussed in a future paper. Here we quantify the kinematic misalignment angle $\Psi$ as the difference between the position angle defined by the photometric major axis (${\rm PA_{phot}}$) and the global kinematic position angle (${\rm PA_{kin}}$) approximately within 1 effective radius. We measure ${\rm PA_{kin}}$ using the method presented in Appendix C of \citet{2006MNRAS.366..787K}\footnote{See footnote~\ref{ft:CapSoft} for software availability.}, which provides a global orientation of the velocity map. ${\rm PA_{phot}}$ was measured  by calculating the moments of inertia\footnote{The routine can be found within the MGE Package \citep{2002MNRAS.333..400C} at http:/www.purl.org/cappellari/} of the surface brightness distribution from the MUSE white-light images (obtained by summing the MUSE cubes along the wavelength dimension). At the same time, the method provides the global ellipticity $\epsilon$. As we used MUSE cubes for both ${\rm PA_{kin}}$ and ${\rm PA_{phot}}$, they were estimated approximately within the same region.  In Table~\ref{t:sample} we report the measured photometric and kinematic position angles as well as other relevant properties used in this paper.

\begin{figure}
\includegraphics[width=\columnwidth]{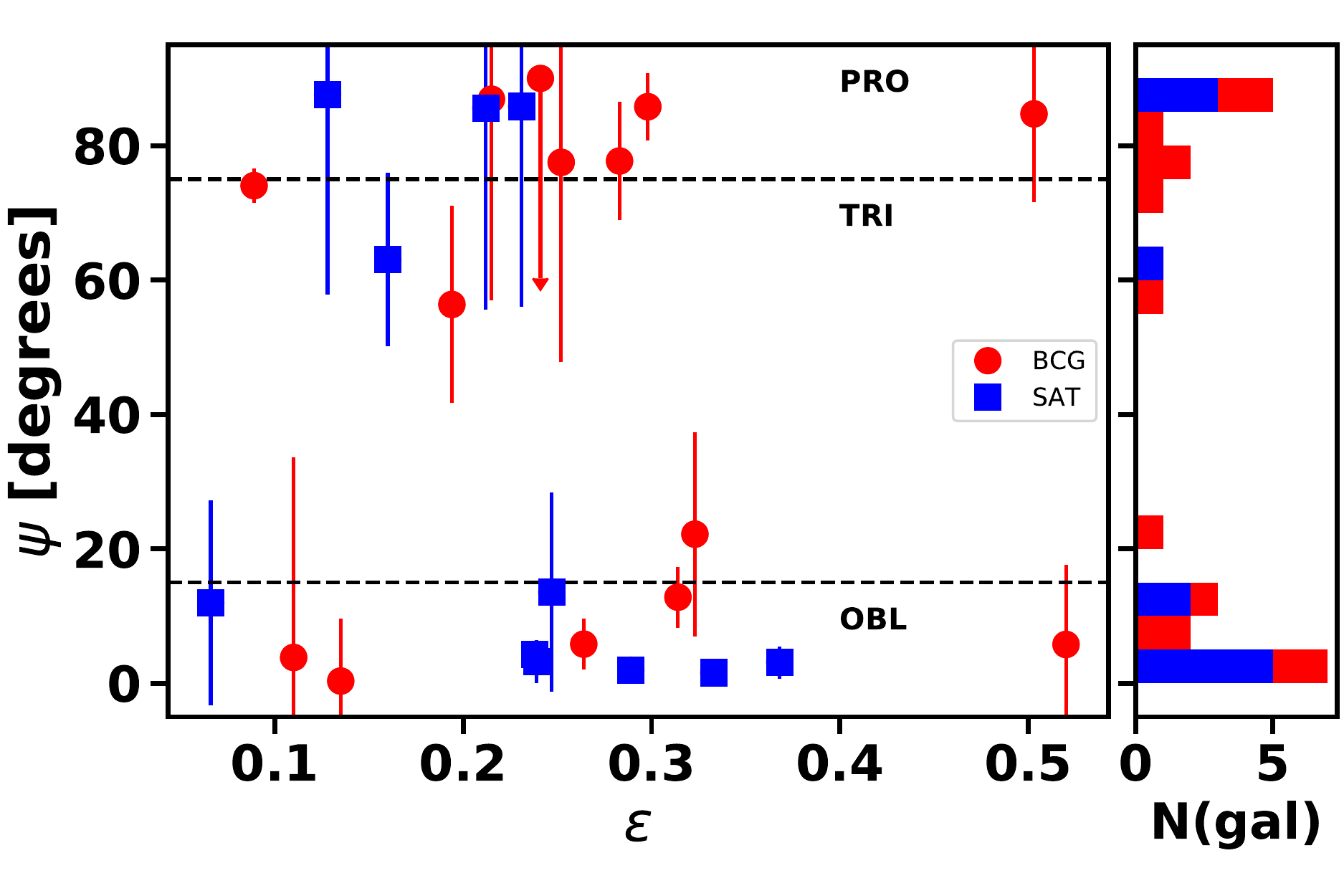}
\caption{Distribution of the kinematic misalignment angle as a function of ellipticity, both measured within the effective radius of M3G sample galaxies. Red circles show BCGs, while blue squares are non-BCGs in the SSC (we call them satellites or SAT for simplicity). The symbol with an upper limit error bar is PGC\,043900, the system with no net rotation and, therefore, no reliable ${\rm PA_{kin}}$ measurement. Horizontal lines at $\Psi = 15$\degr and 75\degr are used to guide the eye for an approximate separation of shapes of galaxies, between mostly oblate (indicated with ``OBL"), triaxial (``TRI") and prolate (``PRO"). These divisions are not meant to be rigorous but indicative. Colours on the right-hand side histogram follow the same convention as shown on the main plot and the legend.}
\label{f:kma}
\end{figure}

Kinematic and photometric position angles are shown in Fig.~\ref{f:maps} as green and brown lines, respectively. Systems with regular rotation have almost overlapping lines, while systems with non-regular rotation often show that the kinematic misalignment angle $\Psi$ is close to 90\degr. To quantify this, we also present the distribution of $\Psi$ as a function of the galaxy projected ellipticity $\epsilon$ for the M3G sample in Fig.~\ref{f:kma}. We split galaxies into BGCs and satellites and draw two horizontal lines at 15\degr~and 75\degr~to separate oblate, triaxial and prolate geometries. 

The most noteworthy characteristic of Fig.~\ref{f:kma} is that galaxies seem to group in two regions, one with low and one with high $\Psi$. Galaxies with $\Psi<15$\degr\, are generally consistent with having oblate symmetries. Their velocity maps look regular, and all galaxies with high rotation amplitudes are found in this group. In order of rising ellipticity these BCGs are: {\it l)} PGC\,004500, {\it m)} PGC\,049940, {\it a)} PGC\,047202, {\it j)} PGC\,018236 and {\it i)} PGC\,019085. Their intrinsic shapes are likely not axisymmetric, as their velocity maps show kinematic twists and are not regular, but the velocity maps are close to aligned with their photometric axes. 

Galaxies with $\Psi$ significantly larger than 0 (and lower than 90) cannot be axisymmetric as their net angular momentum is not aligned with one of the principle axes. Very indicative is also that 8 galaxies have $\Psi>75$\degr, while for one galaxy ({\it e} PGC\,04390) it was not possible to determine $\Psi$ as it does not show rotation. Among those 8 galaxies a closer examination shows rotation {\it around} the major axis within a large fraction of the half-light radius. These galaxies exhibit prolate-like rotation, as it is defined in Section~\ref{s:intro}, within a significant part of the MUSE field-of-view. The rotation amplitude is, as in the case of other non-regular rotators, typically small, mostly around 50 km/s or lower, and the observed (luminosity-weighted) rotation has to be supported by the existence of long-axis tube orbits.

\begin{table*}
   \caption{General properties of the galaxies.}
   \label{t:sample}
\begin{tabular}{ccrrrccccccc}
    \hline
    \noalign{\smallskip}
Name & M$_{\rm K}$& PA$_{\rm kin}$ & PA$_{\rm phot}$ & $\Psi$ & $\epsilon$ & $\sigma_e$ & $R_{e}$ & log (M$_\ast$) & BCG & Fig.~\ref{f:maps}\\
          & 2MASS mag& [degree] & [degree] & [degree] & & [km/s] & [kpc] & [log(M$_\odot$)] & \\
  (1) & (2) & (3) & (4) & (5) & (6) & (7) & (8) & (9) &(10) & (11)\\        
    \noalign{\smallskip} 
    \hline \hline
    \noalign{\smallskip}
PGC003342 & -26.34&139.2 $\pm$ 29.7 & 36.7 $\pm$ 0.3 & 77.5 $\pm$ 29.7 & 0.25 & 270.0 $\pm$ 1.6 & 24.5 & 12.283 & 1 & f)\\
PGC004500 & -25.85&157.3 $\pm$ 29.7 & 153.5 $\pm$ 1.4 & 3.8 $\pm$ 29.7 & 0.11 & 256.5 $\pm$ 1.6 & 13.2 & 12.026 & 1 & l)\\
PGC007748 & -25.76&143.7 $\pm$ 8.8 & 66.0 $\pm$ 0.8 & 77.7 $\pm$ 8.8 & 0.28 & 265.6 $\pm$ 1.2 & 15.9 & 12.072 & 1 & n)\\
PGC015524 & -26.10&116.5 $\pm$ 12.6 & 172.9 $\pm$ 7.5 & 56.4 $\pm$ 14.6 & 0.19 & 272.9 $\pm$ 1.4 & 20.7 & 12.233 & 1 & g) \\
PGC018236 & -25.93&52.9 $\pm$ 4.5 & 65.7 $\pm$ 0.5 & 12.8 $\pm$ 4.5 & 0.31 & 274.5 $\pm$ 1.5 & 18.3 & 12.142 & 1 & j)\\
PGC019085 & -26.04&119.5 $\pm$ 11.8 & 125.3 $\pm$ 0.3 & 5.8 $\pm$ 11.8 & 0.52 & 261.2 $\pm$ 1.4 & 28.4 & 12.358 & 1 & i)\\
PGC043900 & -26.46&146.7 $\pm$ 89.1 & 177.5 $\pm$ 1.7 & $<$ 90 & 0.24 & 376.7 $\pm$ 2.2 & 25.8 & 12.539 & 1 & e)\\
PGC046785 & -26.39&63.5 $\pm$ 4.0 & 75.4 $\pm$ 14.7 & 11.9 $\pm$ 15.2 & 0.07 & 333.9 $\pm$ 1.8 & 18.2 & 12.464 & 3 & o)\\
PGC046832 & -26.59&167.9 $\pm$ 11.8 & 145.7 $\pm$ 9.6 & 22.2 $\pm$ 15.2 & 0.32 & 311.9 $\pm$ 1.4 & 21.4 & 12.359 & 2 & d)\\
PGC046860 & -25.89&56.0 $\pm$ 1.8 & 57.6 $\pm$ 1.0 & 1.6 $\pm$ 2.0 & 0.33 & 282.8 $\pm$ 1.2 & 13.0& 12.038  & 3 & v)\\
PGC047154 & -26.29&71.1 $\pm$ 12.9 & 134.2 $\pm$ 0.2 & 63.1 $\pm$ 12.9 & 0.16 & 319.5 $\pm$ 2.2 & 16.0 & 12.254 & 3 & p)\\
PGC047177 & -25.89&95.3 $\pm$ 2.0 & 98.4 $\pm$ 1.3 & 3.1 $\pm$ 2.4 & 0.37 & 287.3 $\pm$ 1.5 & 15.3 & 12.104 & 3 & u)\\
PGC047197 & -25.99&24.2 $\pm$ 29.7 & 118.4 $\pm$ 2.1 & 85.8 $\pm$ 29.8 & 0.23 & 301.7 $\pm$ 1.8 & 14.3 & 12.136 & 3 & t)\\
PGC047202 & -27.19&166.4 $\pm$ 2.8 & 172.2 $\pm$ 2.5 & 5.8 $\pm$ 3.8 & 0.26 & 318.9 $\pm$ 2.2 & 39.3 & 12.668 & 2 & a)\\
PGC047273 & -25.69&96.8 $\pm$ 2.3 & 93.6 $\pm$ 2.2 & 3.2 $\pm$ 3.2 & 0.24 & 261.3 $\pm$ 1.4 & 18.4 & 12.064 & 3 & y)\\
PGC047355 & -25.79&86.2 $\pm$ 1.0 & 81.9 $\pm$ 1.2 & 4.3 $\pm$ 1.6 & 0.24 & 252.9 $\pm$ 1.3 & 14.8 & 12.026 & 3 & x)\\
PGC047590 & -26.09&46.9 $\pm$ 29.7 & 132.4 $\pm$ 3.8 & 85.5 $\pm$ 29.9 & 0.21 & 288.8 $\pm$ 2.1 & 16.8 & 12.254 & 3 & r)\\
PGC047752 & -25.89&3.0 $\pm$ 29.7 & 89.9 $\pm$ 3.3 & 86.9 $\pm$ 29.9 & 0.22 & 261.0 $\pm$ 1.7 & 21.2 & 12.073 & 2 & k)\\
PGC048896 & -26.70&87.7 $\pm$ 13.1 & 3.0 $\pm$ 0.5 & 84.7 $\pm$ 13.1 & 0.50 & 322.0 $\pm$ 1.9 & 34.9 & 12.611 & 1 & b) \\
PGC049940 & -25.83&51.4 $\pm$ 9.1 & 51.7 $\pm$ 1.9 & 0.3 $\pm$ 9.3 & 0.14 & 298.6 $\pm$ 1.7 & 12.6 & 12.078 & 1 & m)\\
PGC065588 & -26.05&136.1 $\pm$ 2.3 & 62.1 $\pm$ 1.1 & 74.0 $\pm$ 2.5 & 0.09 & 274.0 $\pm$ 2.1 & 22.3 & 12.309 & 1 & h)\\
PGC073000 & -26.65&66.6 $\pm$ 5.0 & 160.8 $\pm$ 0.3 & 85.8 $\pm$ 5.0 & 0.30 & 283.6 $\pm$ 1.8 & 30.4 & 12.450 & 1 & c)\\
PGC097958 & -25.79&45.4 $\pm$ 1.5 & 47.3 $\pm$ 0.3 & 1.9 $\pm$ 1.5 & 0.29 & 285.9 $\pm$ 1.3 & 9.70 & 11.972 & 3 & w)\\
PGC099188 & -25.99&149.7 $\pm$ 29.7 & 62.1 $\pm$ 1.4 & 87.6 $\pm$ 29.7 & 0.13 & 224.2 $\pm$ 1.2 & 26.2 & 12.198  & 3 & s)\\
PGC099522 & -26.09&102.9 $\pm$ 11.1 & 89.4 $\pm$ 9.8 & 13.5 $\pm$ 14.8 & 0.25 & 234.6 $\pm$ 0.9 & 19.0 & 12.036  & 3 & q)\\

     \noalign{\smallskip}
    \hline
 \end{tabular}
\\
{Notes: Column 1: names of galaxies; Column 2: Absolute magnitudes; Column 3: kinematic position angle; Column 4: photometric position angle: Column 5: kinematic misalignment error;  Column 6: ellipticity; Column 7: velocity dispersion within the effective radius; Column 8: effective radius based on the {\tt j\_r\_eff} 2MASS XSC keyword; Column 9: stellar mass; Column 10: galaxy is a BCG - 1, galaxy is a BCG in the SSC - 2, galaxy is a ``satellite" in the SSC - 3; Column 11: the letter referring to the position of the object in Fig.~\ref{f:maps}. Absolute K-band magnitudes are based on the 2MASS K-band total magnitudes and the distance moduli obtained from NED (http://ned.ipac.caltech.edu). The same distance moduli were used to convert sizes to kiloparsecs. Note that while we report actual measurements for the kinematic and photometric position angles, the kinematic misalignment $\Psi$ for PGC043900 is an upper limit, as there is no net streaming in this galaxy. The stellar mass reported in the last column was estimated using columns 7 and 8, and the virial mass estimator from \citet{2006MNRAS.366.1126C}. }
\end{table*}

There are 4 galaxies with $15^{\rm o}<\Psi<75^{\rm o}$: {\it h)} PGC\,065588, {\it p)} PGC\,047154, {\it g)} PGC\,015524 and {\it d)} PGC\,046832 (in order of decreasing $\Psi$). The first three have similar rotation pattern as other galaxies with prolate-like rotation. Strictly speaking the $\Psi$ values are inconsistent with 90\degr, but their velocity maps resemble those of galaxies with prolate-like rotation. We will, therefore, also refer to them as having prolate-like rotation. On the other hand, {\it d)} PGC\,046832 exhibits a very complex velocity map with multiple changes between approaching and receding velocities, but its velocity map does not resemble prolate-like rotation, and its $\Psi$ is significantly smaller than for other galaxies in this group. Therefore, we will not consider it to have prolate-like rotation. As mentioned before, another special case is {\it e)} PGC\,043900, which does not have any rotation and its $\Psi$ is not well defined. Therefore, it is plotted as an upper limit. 

The prolate-like rotation comes in two flavours. It can be present across the full observed field-of-view (approximately 2 effective radii), for example in {\it n)} PGC007748, {\it h)} PGC065588 and {\it r)} PGC047590, but most galaxies have it within a specific area, either outside the central region (but within one effective radius, {\it s)} PGC099188), or more typically covering the full half-light radius (e.g. {\it k)} PGC047752, {\it f)} PGC003342, {\it c)} PGC007300 or {\it b)} PGC048896). In these cases, the rotation at larger radii either disappears (e.g. {\it t)} PGC047197) or there is a change in the kinematic position angle and the rotation is consistent with being around the minor axis ({\it f)} PGC003342, {\it c)} PGC073000, {\it b)} PGC048896). The change in the kinematic position angle is relatively abrupt and occurs over a small radial range. Therefore, such galaxies could even be characterised as having large-scale KDCs, with the central component exhibiting prolate-like rotation. More typical and standard size KDCs are found in a few M3G targets ({\it i)} PGC019085 and {\it d)} PGC046832), but these will be discussed in more detail in a future paper devoted to the analysis of the high spatial resolution MUSE data cubes. 

Finally, for a few galaxies there is evidence for a significant change beyond one effective radius in the properties of the velocity maps: regardless of regular or non-regular rotation within the effective radius, the outer parts show no rotation. They are, however, characterised by a spatially symmetric shift of velocities to larger values compared to the systemic velocity of the galaxy. Examples are the BCGs: {\it m)} PGC049940, {\it i)} PGC019085 and {\it g)} PGC015524. Except for stressing that such velocities at larger radii are only found in the BCGs, we will postpone the discussion of these features to a future paper when it will be put in the full context of the kinematics of M3G galaxies.

%
%
\section{Discussion}
\label{s:discs}

\begin{figure}
\includegraphics[width=0.5\textwidth]{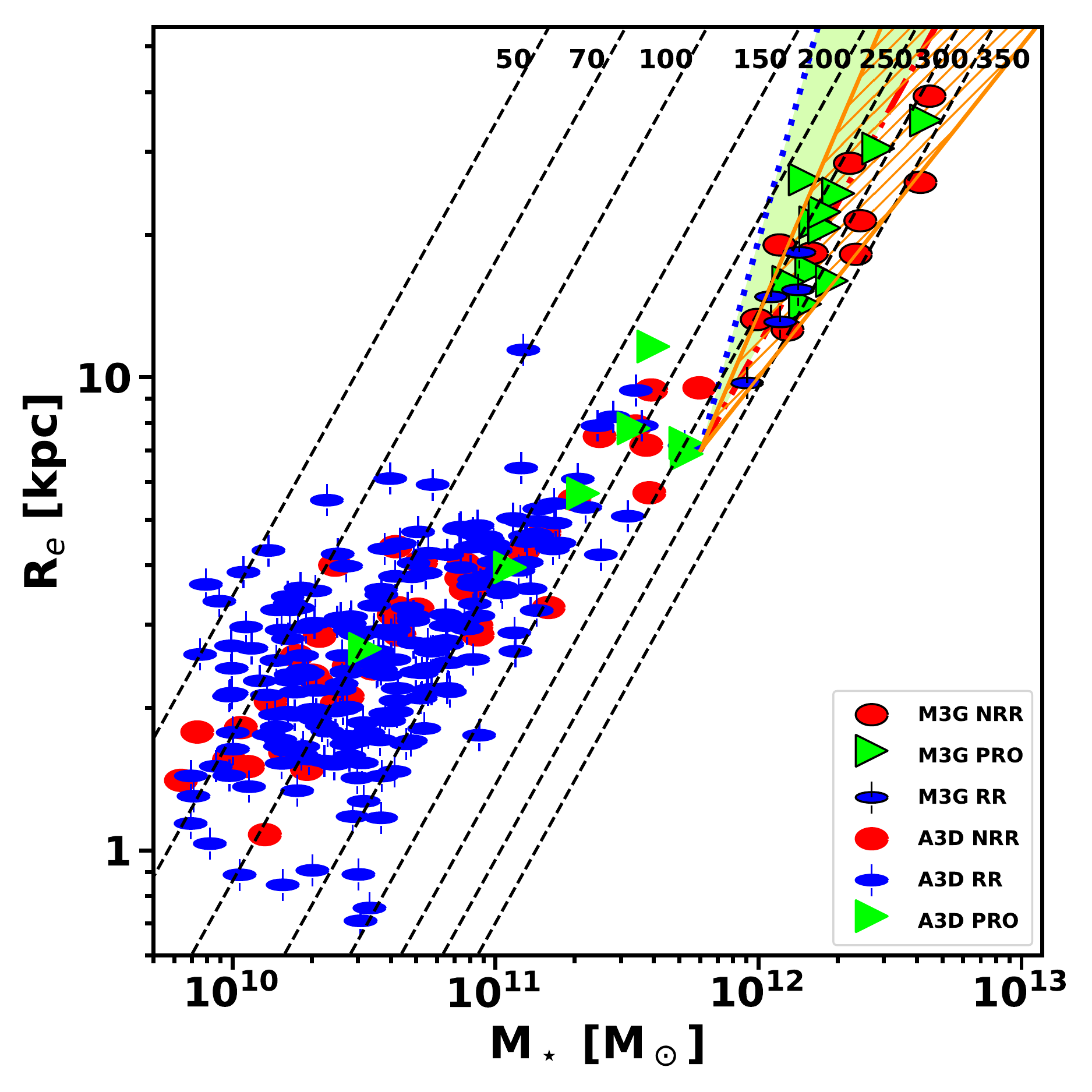}
\caption{The distribution of the M3G sample on the mass size plane. The M3G sample is shown with symbols that have black edges and dominate the high-mass end. For reference we also show galaxies from the ATLAS$^{\rm 3D}$ sample with coloured symbols. The shape and the colour of the symbol is related to the kinematic type as indicated in the legend. The classification is taken from \citet{2011MNRAS.414.2923K} with the following meanings: RR - regular rotation, NRR - non-regular rotation, and PRO - prolate-like rotation (nominally the latter are part of the NRR group, but we highlight them here). Diagonal dashed lines are lines of constant velocity dispersion calculated using the virial mass estimator. The green shaded region shows the expected region where galaxies growing through dissipation-less mergers should lie, assuming major 1:1 mergers (dot-dashed red line) and multiple minor mergers (dotted blue line). The orange hatched region encompasses the mass - size evolution of major merger remnants depending on the merger orbital parameters, as explained in Section~\ref{s:discs}.}
\label{f:ms}
\end{figure}

In Fig.~\ref{f:ms} we place the M3G sample on the mass - size diagram. We indicate the type of observed kinematics with different symbols and colours and also add the galaxies from the ATLAS$^{3D}$ magnitude-limited sample for comparison. Galaxy masses and sizes for ATLAS$^{3D}$ galaxies were obtained from \citet{2013MNRAS.432.1709C}. For M3G objects we used their 2MASS sizes (XSC keyword {\tt j\_r\_eff}), defining the size as $R_e = 1.61\times \rm ${\tt j\_r\_eff} as in \citet{2013ApJ...778L...2C}. Masses of the sample galaxies were approximated using the virial mass estimator M$_\ast = 5\times(R_e\sigma_e^2$)/G \citep{2006MNRAS.366.1126C}, where $\sigma_e$ is the effective velocity dispersion extracted from the MUSE data within an ellipse of area equal to $\pi\times R_e^2$. 

Using the full M3G sample, up to 44 per cent of galaxies have prolate-like rotation (here we include {\it h)} PGC\,065588, {\it p)} PGC\,047154 and {\it g)} PGC\,015524 with $\Psi>60$\degr, but do not consider {\it e)} PGC\,043900). The M3G objects located in the SSC form a magnitude-limited subsample within a well defined environment. This subsample contains 5/14 (35 per cent) galaxies with prolate-like rotation. Of these 5 galaxies one is a BCG, while the other two BCGs in the SSC, including the most luminous and the largest galaxy in the sample, do not show prolate rotation. The fraction of prolate-like rotation is somewhat higher among BCGs. In our sample there are 7/14 BCGs with prolate-like rotation (excluding {\it e)} PGC\,043900 with an uncertain $\Psi$), or 50 per cent. A comparison with the ATLAS$^{3D}$ sample indicates that galaxies with prolate-like rotation are mostly found in massive galaxies and that they are typical for dense environments. This can be quantified using the literature data. 

Within the ATLAS$^{\rm 3D}$ sample there are six known galaxies with prolate-like rotation (NGC\,4261, NGC\,4365, NGC\,4406, NGC\,5485, NGC\,5557 and NGC\,4486), while \citet{2017A&A...606A..62T}  found 8 new systems in the CALIFA sample \citep[LSBCF560-04, NGC0810, NGC2484, NGC4874, NGC5216, NGC6173, NGC6338, and UGC10695; ][]{2017A&A...597A..48F}\footnote{We excluded NGC\,5485 as it is already in ATLAS$^{3D}$ sample. UGC10695 is only a candidate for prolate-like rotation.}. Together with the previously known cases such as NGC1052 \citep{1979ApJ...229..472S},  NGC4589, NGC5982 and NGC7052 \citep{1988A&A...195L...5W}, this means a total of 17 galaxies with apparent prolate-like rotation were previously known in the nearby universe. The MASSIVE survey \citep{2014ApJ...795..158M} found 11 galaxies with kinematic misalignment larger than 60\degr, whereas 7 of those have $\Psi>75$\degr\, and can therefore be considered to have prolate-like rotation \citep{2018arXiv180200014E}. These galaxies are: NGC\,708, NGC\,1060, NGC\,2783, NGC\,2832, NGC\,7265, NGC\,7274, and UGC\,2783, where all of them except NGC7274 are classified as BCGs or brightest group galaxy (BGG). A recent study of the kinematic misalignment angle of more than 2000 MANGA galaxies \citep{2018MNRAS.tmp..522G} finds also a secondary peak at $\Psi\sim90$\degr\, among galaxies more massive than $2\times10^{11}$ M$_\odot$. Combining the M3G sample of galaxies with prolate-like rotation with those from the literature, we see that such rotation typically does not occur for M$_\ast \lesssim 10^{11}$ M$_\odot$, and that for M$_\ast \gtrsim 10^{12}$ M$_\odot$ velocity maps with prolate-like rotation correspond to the most populated kinematic category.

Within the M3G sample, the prolate-like rotation is mostly found in BCGs, but is also present in non-BCGs. However, all galaxies in the M3G sample are members of groups or clusters of galaxies. Even when including the literature data, most galaxies with prolate-like rotation have been observed in galaxy clusters or groups. A similar finding is reported by the MASSIVE survey \citep{2018arXiv180200014E}, where galaxies with prolate-like rotation are almost exclusively found in BCGs/BGGs, and generally misaligned galaxies ($\Psi>15$\degr) are rare in the low density environments, but common among the BCGs/BGGs or satellites. As the creation of non-regularly rotating, massive galaxies with low angular momentum (typical hosts for prolate-like rotation) can a priori occur in any environment \citep[e.g.][]{2011MNRAS.416.1680C, 2017arXiv170308573V}, we expect that galaxies with prolate-like rotation, if rare, still exist outside of dense environments. The evidence that this might be so could be seen in recent merger galaxies, such as NGC\,1222 \citep{2018MNRAS.tmp..513Y} or NGC\,7252 \citep{2018arXiv180109691W}. These galaxies are in late merging phases, and have not yet fully settled, but show prolate-like rotation of the stellar component. What makes them significantly different from other prolate-like systems, is their richness in atomic and emission-line gas, as well as ongoing star formation, implying that there are multiple ways of creating prolate-like kinematics. Such galaxies seem however rare, as \citet{2015A&A...582A..21B} does not report a significant incidence of large kinematic misalignment in mergers. A survey of massive galaxies across various environments could constrain the dependence of prolate-like rotation on the environment, as well as offer new possible scenarios for their formation. 

Numerical simulations suggest that prolate-like rotation may be the outcome of binary mergers for specific orbital configurations \citep[e.g][]{2014MNRAS.445L...6L}. For example, major (1:1) dissipation-less mergers in the study by \citet{2003ApJ...597..893N} exhibit rotation around the minor axis. Furthermore, the orbital structure and the shapes of remnants of major collisionless mergers indicate significant triaxiality and dominance of orbits that support triaxial or prolate shapes \citep{2005MNRAS.360.1185J,2009MNRAS.397.1202J,2014MNRAS.445.1065R}. Numerical simulations of binary (disk) mergers often end up with mildly elongated and low angular momentum remnants, with triaxial shapes and prolate-like rotation \citep{1992ApJ...400..460H, 2003ApJ...597..893N,2006ApJ...650..791C,2010ApJ...723..818H,2011MNRAS.416.1654B, 2014MNRAS.444.1475M}. 
More specifically, \citet{2017A&A...606A..62T} emphasised that a polar merger of gas-free disc galaxies can lead to a prolate-like remnant. \citet{2015ApJ...813...10E}, looking at a broader set of merging configurations, found that radial orbits are more likely to produce prolate-like rotation, other orbital configurations (specific combinations of orbital and disk angular momentum) not being excluded.  

Similar results are recovered in numerical simulations set within a cosmological context. Cosmological zoom-in simulations produce galaxies with prolate-like rotation \citep{2014MNRAS.444.3357N}. The Illustris \citep{2014MNRAS.444.1518V}, EAGLE \citep{2015MNRAS.446..521S} and cosmo-OWLS \citep{2014MNRAS.441.1270L} numerical simulations find that there is an increasing fraction of (close to) prolate shapes among the most massive galaxies \citep[][for EAGLE+cosmo-OWLs and Illustris, respectively]{2015MNRAS.453..721V, 2016MNRAS.455.3680L}. Major mergers seem to be ubiquitous among galaxies with prolate-like rotation \citep{2017arXiv170803311E}. Specifically, a late (almost dry) major merger seems to be crucial to decrease the overall angular momentum and imprints the prolate-like rotation. A recent study by \citet{2018MNRAS.473.1489L} on the origin of prolate galaxies in the Illustris simulation, shows that they are formed by late ($z < 1$) major dissipation-less mergers: galaxies might have a number of minor or intermediate mass mergers, but the last and late major merger is the main trigger for the prolate shape. Similarly to the findings from idealised binary mergers, most mergers leading to prolate-like systems have radially biased orbital configurations. Lower-mass remnants may allow a broader set of possible orbital parameters, mass ratios as well as gas content among the (higher angular momentum) progenitors leading to prolate-like rotation \citep{2017arXiv170803311E}.

Prolate-like rotation does not strictly imply that the galaxy has a prolate mass distribution (or potential). This is nicely illustrated with idealised St\"ackel potentials, where prolate systems allow only inner and outer long-axis tube orbits \citep{1985MNRAS.216..273D}. Hence prolate galaxies can have velocity maps that either show prolate-like rotation or no-rotation. This is indeed found for the Illustris prolate-like galaxies; about 51\% of actually prolate galaxies (using a tri-dimensional account of the mass distribution) show prolate-like rotation while the others have no net rotation \citep{2018MNRAS.473.1489L}, presumably as they contain both prograde and retrograde long-axis tube orbits. Nevertheless, galaxies with prolate-like rotation cannot be oblate spheroids. 

Velocity maps of the M3G sample objects with prolate-like rotation show spatial variations, sometimes changing at larger radii to rotation around the major axis. This suggests more complex shapes, supporting various types of orbital families \citep{1991ARA&A..29..239D}. A classical example of such galaxies is NGC\,4365 \citep{1988A&A...202L...5B}, which has a large KDC and outer prolate-like rotation. Its orbital distribution is complex with both short- and long-axis tubes responsible for the formation of the observed (luminosity-weighted) kinematics \citep{2008MNRAS.385..647V}. This is also a characteristic of high-mass merger remnants, which often contain a large fraction of box orbits, short- and long-axis tubes, varying relatively to each other with radius \citep[e.g.][]{2014MNRAS.445.1065R}. With such caveats in mind, it is worth assuming for a moment that M3G galaxies with prolate-like rotation are actually significantly triaxial and close to being prolate. Prolate galaxies in the Illustris simulation are found only at masses larger than $3\times10^{11}$ M$_\odot$, and above 10$^{12}$ M$_\odot$ 62 per cent of galaxies are prolate or triaxial, 43 per cent being prolate \citep{2018MNRAS.473.1489L}. This is coincidentally close to our observed fraction of prolate-like systems (44\%) within the M3G sample. The similarity between these fractions should be taken with caution, as we stress the M3G sample is neither complete nor a representative sample, and the number of actually prolate galaxies is certainly lower then the number of galaxies with prolate-like rotation.

Notwithstanding the actual frequency and shape of galaxies with prolate-like rotation, they cluster in a special region of the mass - size diagram, as Fig.~\ref{f:ms} shows. The M3G sample lies on an extension of the arm-like protuberance arising from the cloud of galaxies at high masses and large sizes. The M3G data extend this arm by almost an order of magnitude in mass and a factor of 5 in size. At masses below $6\times10^{11}$ M$_\odot$ covered by previous surveys, galaxies that were found on this extension were typically old and metal-rich slow rotators characterised by a deficit of light (cores) in their nuclear surface brightness profiles \citep{2011MNRAS.414..888E,2013MNRAS.432.1862C, 2013MNRAS.433.2812K, 2015MNRAS.448.3484M}. Specifically, their kinematic properties and the core-like light profiles were used as an indication that the formation of these galaxies was different from other galaxies populating the mass - size plane, which are characterised as star-forming disks, or bulge dominated, oblate and fast rotating early-type galaxies \citep{2013MNRAS.432.1862C, 2016ARA&A..54..597C}.  The most likely formation process of galaxies populating that extension is through dissipation-less mergers of already massive galaxies: these may provide a way to explain their kinematics, low angular momentum content, cores in light profiles \citep[through binary black hole mergers, e.g.][]{1991Natur.354..212E,2001ApJ...563...34M} and old stellar populations.

The M3G extension of the arm supports this picture in two additional ways. Firstly, it shows that while these galaxies span a large range in both mass and size, their effective velocity dispersions are not very different, as expected in major dissipation-less mergers \citep[e.g.][]{2009ApJ...691.1424H,2009ApJ...697.1290B,2009ApJ...699L.178N}. Following the argument outlined in \citet{2009ApJ...699L.178N}, if a massive galaxy grows via equal-mass mergers (of progenitors with similar sizes and/or velocity dispersions), both the mass and the size of the remnant will increase by a factor of 2, while it will follow a line of constant velocity dispersion in Fig.~\ref{f:ms}. We illustrate this path with a red dot-dashed line, where products of consecutive equal-mass mergers would fall, for example starting with systems of M$ =6\times10^{11}$ M$_\odot$ and R$_{e}=7$ kpc, representative of the most massive galaxies in the local Universe. The same increase in mass achieved through multiple minor mergers (with smaller mass, size and velocity dispersion progenitors) would lead to a size increase by a factor of 4, while the velocity dispersions would typically be reduced by a factor of 2. This corresponds to the blue dotted line in Fig.~\ref{f:ms}, starting from the same main galaxy progenitor \citep[see also fig. 2 in][]{2009ApJ...697.1290B}.

Equal mass merger simulations show that the relation between the mass and the size of galaxies also depends on the merger parameters, such as the pericentric distance, type of the orbit and its angular momentum \citep{2006MNRAS.369.1081B}. This study showed that depending on the merger orbit, the mass - size relations follows $R_e \ \sim M_\ast^\alpha$, where $\alpha=0.7-1.3$. We add this range of possibilities on Fig.~\ref{f:ms} as a hatched region, indicating the possible location for massive galaxies after major mergers, and fully encompassing M3G sample galaxies. A caveat in this simple argument is that some of the massive galaxies today will start merging as more compact objects in the early Universe, as is evident from the evolution of the mass - size relation with redshift \citep{2014ApJ...788...28V} and implied by compact size of high redshift quiescent galaxies and their subsequent evolution \citep[e.g.][]{2008ApJ...677L...5V,2010ApJ...709.1018V}.

Inevitably the merger history of massive galaxies will be a combination of multiple minor mergers and a small number of major (or even equal) mass mergers \citep[e.g][]{2007MNRAS.375....2D,2012ApJ...754..115J,2014MNRAS.444.3357N}. The evidence for such a combination is visible in the differences between the central region (about 1 R$_e$) and the outskirts, as they often do not share the same kinematics or stellar populations, which will be the topic of future papers. The tightness of the region on the mass - size diagram within which the M3G galaxies lie suggests that the growth of the most massive galaxies ($>10^{12}$ M$_\odot$) and, in particular, BCGs is dominated by major mergers. This would be consistent with the findings by \citet[]{2018MNRAS.473.1489L} that also links such massive mergers with prolate-like rotation. Given that more than half of the BCGs in our sample exhibits prolate-like rotation, we speculate that indeed most of these experienced a late major (dry) merger, between two massive (possibly both central) galaxies. A radial bias in the orbital configurations for such mergers leading to an increase fraction of prolate-like rotators may naturally emerge from the preset of phase-space distribution of massive galaxies, also relative to the large-scale structures \citep{1995ApJ...451L...5W, 2010MNRAS.405.2023N, 2017NatAs...1E.157W}.

\section{Conclusions}
\label{s:con}

In this work, we report that a large fraction of galaxies more massive than $10^{12}$ M$_\odot$ show prolate-like rotation. This is shown by the analysis of MUSE data of the magnitude-limited sample of massive galaxies in the Shapley Super Cluster and a matching (in luminosity) sample of BCGs. This M3G sample consists of 25 galaxies, of which 14 are BCGs, 11 are satellites in the SSC and 3 are BCGs in the SSC. We present their stellar velocity maps, and measure their kinematic misalignment angles, showing that 44 per cent of galaxies in the M3G sample have their main rotation {\it around} their major axes. Selecting only BCGs the fraction increases to 50 per cent, while in a magnitude limited subsample of satellites, prolate-like rotation is detected in 35 per cent of galaxies. 

The prolate-like rotation is suggestive of a triaxial or close to prolate intrinsic shape. For most of our galaxies rotation amplitudes are low, but velocity maps typically shows net streaming. These kinematics indicate a violent assembly history, with at least one major dissipation-less merger. The M3G data support a scenario where the final growth of the most massive galaxies is dominated by late dissipation-less merging of similar mass systems. This could be associated with the prevalence of prolate-like rotation in the most massive BCGs and is consistent with the location of these systems within a mass - size diagram, which we extend by almost an order of magnitude in mass and a factor of 5 in size. 

The current sample suggests that there is a rather narrow path for climbing the last rung of the galaxy mass ladder, which would be characteristic of dense cluster environments. Answering whether or not such very massive systems require the merging of already central systems would require a more extended studies and a closer look at relevant simulations. The fact that BCGs seem to show an alignment trend with respect to the larger-scale structures may be a interesting avenue to consider, as it would naturally explain a bias in the orbital configuration for equal-mass massive and late mergers. Interestingly enough, prolate-like rotation is also found in lower-mass galaxies (e.g. as seen by the ATLAS$^{\rm 3D}$ and the CALIFA surveys, as well in some dwarf galaxies). This further suggests that galaxies with prolate-like rotation should be present in low galactic density regions, while the progenitors may be quite different (i.e. gas-rich).

\section*{Acknowledgements}
We wish to thank Joop Schaye, Michael Maseda and Marijn Franx for useful discussions and the full MUSE GTO team for support during observations and preliminary work on this paper. We thank Chung-Pei Ma and the MASSIVE team for sharing their results on the kinematic misalignment. This research has made use of the NASA/IPAC Extragalactic Database (NED) which is operated by the Jet Propulsion Laboratory, California Institute of Technology, under contract with the National Aeronautics and Space Administration. This research made use of Astropy, a community-developed core Python package for Astronomy (Astropy Collaboration, 2013).




\bibliographystyle{mn2e}






\bsp	
\label{lastpage}
\end{document}